\documentclass[11pt]{article}
\usepackage{fullpage}
\usepackage{url}
\usepackage{booktabs}
\usepackage{graphicx}
\usepackage{multirow}
\usepackage{xcolor}

\usepackage[utf8]{inputenc} 
\usepackage[T1]{fontenc}    
\usepackage{hyperref}       
\usepackage{url}            
\usepackage{booktabs}       
\usepackage{amsfonts}       
\usepackage{nicefrac}       
\usepackage{microtype}      
\usepackage{multirow}
\usepackage{color}
\usepackage{amsmath,amssymb} 
\usepackage{pgfplots}
\pgfplotsset{compat=newest}
\usepackage{pdflscape}
\usepackage{graphicx}
\usepackage[export]{adjustbox}
\usepackage{algorithm}
\usepackage{mathtools}
\usepackage{algorithmic}
\usepackage{wrapfig,booktabs}

\title{A Roadmap for Greater Public Use of Privacy-Sensitive Government Data:  Workshop Report}
\author{Chris Clifton (Purdue U. Computer Science) \and
  Bradley Malin (Vanderbilt U.)  \and
  Anna Oganian (CDC / National Center for Health Statistics) \and
  Ramesh Raskar (MIT Media Lab) \and
  Vivek Sharma (MIT Media Lab) \and
  \emph{Additional Authors} \and
  Weiyi Xia (Vanderbilt University Medical Center) \and
  Jeremy Seeman (Penn State University) \and
  Zhiyu Wan (Vanderbilt University Medical Center)\and
  Abhishek Singh (MIT Media Lab) }
\date{7$th$ July, 2021}

\begin{document}
\begin{center}
\large{\textbf{A Roadmap for Greater Public Use of Privacy-Sensitive Government Data:  Workshop Report}}\\
Chris Clifton (Purdue University)\\
  Bradley Malin (Vanderbilt University Medical Center)\\
  Anna Oganian (National Center for Health Statistics, CDC)\\
  Ramesh Raskar (MIT Media Lab)\\
  Vivek Sharma (MIT Media Lab) \\ 
  \emph{ } \\
  \emph{Additional Authors}\\
  Weiyi Xia (Vanderbilt University Medical Center) \\
  Jeremy Seeman (Penn State University) \\
  Zhiyu Wan (Vanderbilt University Medical Center)\\
  Abhishek Singh (MIT Media Lab) 
\end{center}

\section*{Executive Summary}
Government agencies collect and manage a wide range of ever-growing datasets. While such data has the potential to support research and evidence-based policy making, there are concerns that the dissemination of such data could infringe upon the privacy of the individuals (or organizations) from whom such data was collected. To appraise the current state of data sharing, as well as learn about opportunities for stimulating such sharing at a faster pace, a virtual workshop was held on May 21\textsuperscript{st} and 26\textsuperscript{th}, 2021, sponsored by the National Science Foundation and National Institute of Standards and Technologies,
and the White House Office of Science and Technology Policy,
where a multinational collection of researchers and practitioners were brought together to discuss their experiences and learn about recently developed technologies for managing privacy while sharing data. 
The workshop specifically focused on challenges and successes in government data sharing at various levels. 
The first day focused on successful examples of new technology applied to sharing of public data, including formal privacy techniques, synthetic data, and cryptographic approaches.
Day two emphasized brainstorming sessions on some of the challenges and directions to address them.
These included:
\begin{itemize}
    \item Stakeholders
    \item Methods
    \item Equity
    \item Problems and Challenges
    \item Research Agendas
    \item Equity and Unspoken Issues
\end{itemize}

\subsection*{Actionable Items}
While the workshop did not arrive at a formal set of recommendations, the authors of the report captured certain key actionable items from the discussions.  They are summarized as follows:
\begin{itemize}
    \item Build a community by organizing a series of conferences and workshops where researchers and practitioners can participate, discuss the problems, and demonstrate how to deploy solutions. 
    \item Applied and foundational research and development. 
    \item Develop common repositories with a variety of data sharing use cases to stimulate and advance research, \textit{e.g.} the Differential Privacy Synthetic Data Challenge~\cite{NISTDPSyn}.
    \item Engage with stakeholders by educating people from a variety of disciplines so that they understand the privacy issues, the data-sharing issues, and the data driven technologies they use.
    \item Communicate the risks, benefits, challenges, limitations, and constraints of the technologies to help build a community around particular use cases or classes of use cases to develop means to address those issues.
\end{itemize}

\section{Introduction}
Government agencies collect and manage voluminous datasets, which hold great potential to support research, economic development, and evidence-based policy making. The rise of data science and Artificial Intelligence (AI)-based technologies has created new, and yet to be recognized, opportunities for use of this data. Unfortunately, much of this data has the potential to be sensitive, which limits public access.

Data sensitivity arises in various ways.  Several of the obvious ones include:
\begin{itemize}
    \item Individual privacy.  Data may reflect confidential information about people.  Even if the data does not pose significant privacy risks, when combined with other data sources it may result in substantial risks of misuse.
    
    \item Corporate confidentiality.  Data may contain important business information that is not public, and its release could have substantial economic impact on the entities the data are about.
    
    \item Political sensitivity.  Datasets held by agencies may contain information whose disclosure could be misinterpreted.  For example, data on individuals used by the Department of Justice for statistical analysis could be misconstrued as surveillance when viewed out of context.
\end{itemize}

There are also less obvious, but still significant, impediments to sharing data widely. These include 1) regulatory and policy-based restrictions, 2) technological impediments, such as how the data are formatted and stored, and 3) budget-imposed constraints, such as lack of public-facing metadata and documentation on the source, quality, and semantics of data.

The Federal government has a long history of data collection. In the United States, the collection of potentially sensitive information on the citizenry dates to the Constitution (while not specifically required by the constitution, the first decennial census asked about gender, race, relationships, slaveholder status, and in some states, occupation)~\cite{CensusHistory} and the requirement that some government collected information be shared at least up until the Patent Act of 1790~\cite{Patent1790}.  The 20\textsuperscript{th} century saw growing concern with privacy of that data covered by the Patent Act of 1790, perhaps most famously through the 1890 article ``The Right to Privacy''~\cite{WarrenBrandeis}, leading to a patchwork of laws, regulations, and policy to restrict access to and prevent the misuse of government held data.

The 21\textsuperscript{st} century has seen a growing awareness of the need to better utilize this data to support public safety, leading to efforts such as the Terrorism Information Awareness program that was established after the 9/11 attack and the 1/21/2021 Executive Order on Ensuring a Data-Driven Response to COVID-19 and Future High-Consequence Public Health Threats.  At the same time, there has been growing concern that such programs, even using already collected data shared only within the government, could be misused leading to substantial public harm. The rise in data breach incidents (and the recognition that government is not immune) has made clear the external risk of misuse of the data; more comprehensive datasets increase the potential damage of such a breach. As a result, efforts to improve data sharing and use have proven challenging and lagged behind private enterprise creation of large and comprehensive (and largely unregulated or lightly regulated) datasets.

Existing fora focus on narrow aspects of these problems, such as technological or policy approaches to data privacy, or user groups of specific types of data. The goal of this workshop was to focus more broadly on impediments to broader use of government data. To accomplish this goal, we aimed to bring together data stewards from across the Federal government with researchers in policy and technology to identify promising directions and open challenges in addressing these impediments.

\section{Workshop}

A two-day virtual workshop was held on May 21\textsuperscript{st} and 26\textsuperscript{th}, 2021, sponsored by the National Science Foundation~(NSF) and the National Institute of Standards and Technologies~(NIST), 
with 40-50 invited participants, primarily drawn from Federal Government agencies. The workshop also included academics with expertise in a variety of policy and technology areas, with an emphasis on those who have demonstrated success in achieving novel data sharing approaches.  

The first day (May 21\textsuperscript{st}) focused on use cases, both successful and unsuccessful, to illustrate both some of the challenges and research ideas that can address those challenges.  Table~\ref{tab:sch1} provides a list of the talks.
\begin{table*}
\caption{Day 1 Schedule May 21st, 2021. \label{tab:sch1}}
\tabcolsep=0.1cm
\centering
\begin{tabular}{l|l}
\toprule
Time (ET) & Schedule \\
\toprule
\multirow{3}{*}{11:00 -11:30} &	Opening Remarks.  \\
&\textbf{Erwin Gianchandani}, National Science Foundation; \\
&\textbf{Lynne Parker} White House Office of Science and Technology Policy \\
\midrule
\multirow{2}{*}{11:30 - 12:00} &	Keynote: Confidentiality-Utility Tradeoffs and Stakeholder Communication. \\ 
& \textbf{John Abowd}, U.S. Census Bureau\\
\midrule
\multirow{2}{*}{12:00 - 12:30} &	Differential Privacy in Practice. \\ 
& \textbf{Ashwin Machanavajjhala}, Duke University\\
\midrule
\multirow{2}{*}{12:30 - 13:00} &Some Elements of the Interface of Privacy Protection with Data Quality, Risk and Cost.\\ 
& \textbf{John Eltinge}, U.S. Census Bureau\\
\midrule
13:00 - 13:15 &	Break\\
\midrule
\multirow{2}{*}{13:15 - 13:45} &	Country-scale deployments of cryptographic security and privacy technologies.\\ 
& \textbf{Dan Bogdanov}, Cybernetica\\
\midrule
\multirow{2}{*}{13:45 - 14:15} &	The Perspective of the Chief Data Officer.\\ 
& \textbf{Dan Morgan}, Department of Transportation\\
\midrule
\multirow{2}{*}{14:15 - 14:45} &	COVID-19 Case Data Publication.\\
&\textbf{Brian Lee}, U.S., Centers for Disease Control and Prevention\\
\midrule
\multirow{2}{*}{14:45 - 15:15} &	Synthetic Data Generation.\\
& \textbf{Jerry Reiter}, Duke University\\
\midrule
15:15 - 15:30 &	Break\\
\midrule
\multirow{2}{*}{15:30 - 15:50} &	Secure Multiparty Computation in the Boston Women's Workforce Network.\\
& \textbf{Mayank Varia}, Boston University\\
\midrule
\multirow{2}{*}{15:50 - 16:10} & 	Multi-tier Access and Data User Needs.\\
&\textbf{Barbara Downs}, U.S. Census Bureau\\
\midrule
\multirow{2}{*}{16:10 - 16:30} &	Adversarial Modeling.\\
& \textbf{Bradley Malin}, Vanderbilt University\\
\midrule
\multirow{2}{*}{16:30 - 16:50} &	The TREC Datasets.\\
&\textbf{Ellen Voorhees}, NIST\\
\midrule
16:50 - 17:00 &	Discussion and Planning for Day 2\\
\bottomrule
\end{tabular}
\end{table*}
 
 The second day (May 26\textsuperscript{th}) consisted of breakout group brainstorming sessions and full-session discussions to identify specific challenges and promising directions to address those challenges.
 Breakout topics were determined based on the discussions on Day 1, and are listed in Table~\ref{tab:sch2}.
\begin{table*}
\caption{Day 2 Schedule May 26th, 2021. \label{tab:sch2}}
\tabcolsep=0.1cm
\centering
\begin{tabular}{l|l}
\toprule
Time (ET) & Schedule \\
\toprule
\multirow{4}{*}{11:00 - 11:30}& 	Opening Remarks. \\
&\textbf{Margaret Martonosi}, National Science Foundation; \\
&\textbf{Elham Tabassi}, National Institute of Standards and Technology;\\
&\textbf{Organizers}\\
\midrule
\multirow{4}{*}{11:30 - 13:00}&	\underline{Session 1, Breakout Sessions}\\
&$\bullet$ Stakeholders\\
&$\bullet$ Methods\\
&$\bullet$ Equity\\
\midrule
13:00 - 13:45&	Break\\
\midrule
13:45 - 14:15&	Breakout Status Reports\\
\midrule
14:30 - 15:00&	Panel on Government - Academia Interaction\\
\midrule
15:00 - 15:15&	Break\\
\midrule
\multirow{4}{*}{15:35 - 16:30}&	\underline{Session 2, Breakout Sessions} \\
&$\bullet$ Problems and Challenges\\
&$\bullet$ Research Agendas\\
&$\bullet$ Equity and Unspoken Issues\\
\midrule
16:30 - 17:00 &	Breakout Summaries\\
\bottomrule
\end{tabular}
\end{table*}
 
\section{Challenges} 
Based on the discussion on Day 1~(Table~\ref{tab:sch1}), the workshop identified several unmet needs of data users, issues limiting broader use of government data, and technologies and policies. From this, five specific challenges were identified:
\begin{itemize}
    \item Test-beds for privacy-aware data sharing (see Section~\ref{subsec:testbeds}).
    \item Stakeholder engagement (see Section~\ref{subsec:stakeholders}).
    \item Equity (see Section~\ref{subsec:equity}).
    \item Technology communication (see Section~\ref{subsec:tech}).
    \item Methods for privacy-preserving data sharing (see Section~\ref{subsec:methods}).
\end{itemize}
The second day of the workshop held small group discussions on issues, concerns, and promising directions to address these challenges (see Table~\ref{tab:sch2}: \underline{Session 1, Breakout Sessions} and \underline{Session 2, Breakout Sessions}.)
Those discussions are summarized in the following subsections.

\subsection{Test-beds} \label{subsec:testbeds}
There is an urgent need for realistic evaluation test-beds for privacy-aware data sharing that can provide insight into the pros and cons of different technologies; others may be facing a ``cold start'' problem about how to initiate and transition to the proper deployment of privacy solutions so significant experience with different technologies may be beneficial. Test-beds could be really helpful in this regard.

Test-beds have largely been developed for advancing specific technologies.  However, a challenge for broader use of public data is determining appropriate policy; this demands stakeholder involvement, technology communication, and other issues that do not fit the traditional mold of challenge problems.  Developing test-beds that can support advances in such areas is an interesting open challenge.

\subsubsection{Public test datasets}
Perhaps the simplest approach to testbeds is to make public test datasets available.  These should be representative of real data, but due to the sensitivity of real data, it can be challenging to release such datasets for research use.

Such datasets generally are created with the expectation that they will support ideas of the challenges (e.g., preventing re-identification or attribute disclosure) and the types of analysis to be performed, but these can be left open.  The type of research to be performed need not be pre-defined. For example, a dataset composed of information derived from electronic health records could be used for experiments in privacy methodologies but may also support research in (for example) stakeholder engagement, with surveys, focus groups, or other research methodologies used to elicit sensitivity, concerns, and perceptions of benefits and risks of making data available.

\paragraph{Real data} is often the preferred choice because it is easy to produce and provides compelling use cases.  However, if data sensitivity is a problem, then releasing real data may not be an option.

A second issue with real data is that there are often issues that complicate analysis of the efficacy of methods developed on that data.  For example, real data has missing values, or may reflect a limited or biased population sample.  As such, a ``ground truth'' is not really known; the quality of methods is measured against a presumed truth that may be distorted.

An example of this is the ``Adult Dataset"~\cite{UCIrepository} commonly used in data privacy and machine learning research.  This is taken from public use microdata that is intended to be used as a weighted sample but is typically used in data privacy research as an unweighted random sample, and thus may not be reflective of how the techniques would work on real data.

\paragraph{Synthetic data} can be used when real data are sensitive.  Furthermore, synthetic data can overcome other issues with real data.  The synthetic data can be treated as ``ground truth'', and sampled or adjusted in multiple ways to simulate various challenges appearing in the real data (e.g., simulating missing data or errors in the data, stratified sampling, bias in sample selection or responses, etc.)


\subsubsection{Controlled access}
In this context, access to protected data requires approval to access the data.
Investigators or investigators' institutions who are interested in obtaining the controlled-access data are required to submit a data access request describing the proposed research use of the data. Access to the data may require user certification and training when dealing with human subjects' identifiers containing specific demographic, clinical, and genotypic information. The access to data can be limited for a given duration. Before the expiration, investigators can submit a request for special permission for renewal of their access period. Controlled access can be realized in a number of ways, such as through virtual enclaves (e.g., the All of Us Program:~\cite{all2019all}), National Center for Health Statistics (NCHS') Research Data Center~(NCHS RDC) or physical enclaves (as the Census does through their Federal Statistical Research Data Centers~\cite{levenstein2020role,fsrdc}).


\subsubsection{Community Challenges}
Providing testbeds in the context of specific challenge problems can spur valuable research.  The TReC (Text Retrieval Conferences) is a prime example, a long-running NIST program that spurred significant advances in information retrieval~\cite{trec}.  While these are best suited to advances in methodology, they can serve other purposes.  For example, challenges can provide a testbed for technology communication, evaluating how other stakeholders perceive and understand the challenge and methodologies developed.
There are multiple ways challenge problems can be arranged.

\paragraph{Competitions} have test datasets, clear goals, and a timeline to advance and demonstrate state-of-the-art methods.  An example of use for data sharing is the NIST Differentially Private Synthetic Data Challenge~\cite{NISTDPSyn,ridgeway2020crisis}.  The competitors were provided with a dataset (e.g., 1940 U.S. Census) and were challenged to produce a synthetic dataset that when analyzed would produce similar statistics to the original data; the entries were scored based on the conformance of statistics to the true values, and on code inspection to ensure differential privacy was satisfied. Another example of use for data sharing is the iDASH (integrating Data for Analysis, 'anonymization' and SHaring)~\cite{tang2016protecting,wang2017community} genome privacy and security competition.  iDASH hosts a critical assessment of data privacy and protection competitions to assess the capacity of cryptographic technologies for protecting computation over human genomes in the cloud.

Competitions are useful for advancing the state-of-the-art for well-understood problems, and in the context of the sharing of sensitive government-held data, can be useful for improving methods.  However, some of the other challenges (such as policy-related challenges) seem less amenable to this format.

Particular opportunities could include competitions similar to the NIST challenge above, but with more complex data (e.g., linked data from multiple sources, higher dimension data, or data over time) or different types of data sharing constraints.

\paragraph{Open datasets with specific goals} provide a more flexible approach, as the utility and ``scoring'' do not need to be clearly defined.  One suggested example was analysis of synthetic data from multiple sources.  The challenge would be that individuals must be linked across sources for the data to make sense, but as the data are synthetically created, the concept of linked data about an individual is not well-defined.  Here the specific scoring and even definition of the exact analysis to be performed may be less specific, allowing a wide variety of approaches to be developed.

\subsection{Stakeholder Engagement} \label{subsec:stakeholders}
\subsubsection{Problem identification and coordination} There is a need to identify specific privacy problems faced by each Federal agency/entity; for this, experts within the agencies need to be engaged. There is a need to foster understanding of such privacy problems in a coordinated way. Such coordinated efforts also will help data users see more clearly data sharing as well as integration issues among multiple agencies, and privacy challenges therein. 

\textbf{Stakeholders and championing are key towards adoption} – In many cases, because of higher trust standards expected from government, non-technical issues pose significant challenges (e.g., legal, policy, and political). There is a need for a strategy to engage stakeholders and appropriate mechanisms to establish champions for deployment of privacy solutions for data sharing.  This strategy should include government representatives, but also the individuals to whom the data corresponds, so that a culture of trust and transparency can be cultivated and sustained.

\subsubsection{Stakeholder Identification}
The first step is to identify different types of stakeholders. Generally, stakeholders include data subjects (those the data are about), data users (those data are shared with), data controllers/stewards (those who collect/hold data), mechanism developers, policymakers (i.e., data regulators), and others. The goals of privacy protections include promoting the participation of subjects, ensuring the ethical use of data, and following policy requirements by using appropriate social and technical mechanisms.  The incentive structures during the data life cycle need to be carefully designed and measured because privacy protections may sometimes disincentivize the engagement of users and the participation of subjects. An example is the Federal Communications Commission's (FCC) speed test data collection~\cite{FCCPP}: if users knew that data users other than the FCC had access to the data, the participation rate might decrease. To solve this problem, FCC could make a statement and guarantee that the data users will do so in a privacy-preserving manner.

Another concrete example is NIST's synthetic data set challenges~\cite{NISTDPSyn,ridgeway2020crisis}: Farmers, the data subjects, are withdrawing from data collection (collected by the data stewards) because of privacy concerns.  If data stewards have a modern high-quality data synthesizer, they may minimize subjects' privacy concerns because identification of individuals is very unlikely in such case. Hence, more subjects may participate, and data can be shared freely with data users. There are several mechanisms for data stewards to address potential problems. First, external verification is required to ensure that the synthetic data cannot be used to reconstruct the original data. Second, external verification is required to ensure that the synthetic data maintain good quality for other query workloads. The effectiveness of a synthetic data generator should be properly measured in terms of utility and privacy to justify its adoption while considering the needs of all stakeholders it serves, particularly the data users and the data subjects.

Other stakeholders that should be considered include the people impacted, as well as the beneficiaries of the data use. For example, in the biomedical sector, genetic data impacts many people well beyond the data subjects (e.g., other family members). Another example can be given in the agricultural sector: a company that leases data collectors the equipment also gains benefits from the data use. A special group of stakeholders are the taxpayers. It might be difficult to make relevant information of a research program funded by a Federal agency available to a stakeholder group like this. For example, the 21\textsuperscript{st} Century Cures Act~\cite{law3024law} requires that all systematic investigations with clinical data be publicly disclosed along with a clear description of why such a study benefits society. Although incentive structures are not Federally required, a market-based model might be viable since more data are upheld as a good. However, restrictions such as a privacy budget, if being implemented too soon, might inherently limit the data utility/value.

Nuanced conversations about the value and potential adverse effects can be promising. For example, regarding the Evidence Act~\cite{evidence}, it is worth noting that Congress specifically did not pass the provisions relating to the National Data Linking system.  Also, it is important to think of ``utility'' in a systematic manner.  That is, if the data stewards were to increase the types of uses, response rates may decrease because privacy risk is increased and thus trust is reduced. As a result, the utility could be reduced in that way too.

Stakeholder engagement is essential for the effectiveness of protection methodologies, such as synthetic data generation, because evaluating the utility and suitability of the synthetic data requires engagement with outside users on what they want to study and a spectrum of use cases. With the engagement of downstream users, synthetic data can be improved effectively with validation.

Making data available beyond traditional researchers and institutions can be beneficial, but it can also cause harm in the other direction. Having a diverse set of stakeholders can create opportunities to hear the potential harms. Those diverse voices need to be brought in early and often. Currently, groups will typically call in external voices, as opposed to just having those with well-shaped roles with a seat at the table. Changes in leadership may affect how this gets done.

\subsubsection{Trust and Transparency}
Trust and transparency among stakeholders is important, if not more important than the aforementioned issues, both in terms of how data are collected, as well as how it is used. Both the data steward and the data user should understand the constraints the other is facing in sharing data. Specifically, a data user should make it clear how the data will be used. 
On the other hand, even if data are shared, if the privacy protection methodology is not transparent to the data users, they may choose to not use the data due to concerns over the data quality.  For example, bias, noise, and missing values might be introduced by a certain protection method. To solve this problem, it is recommended the data steward use a test bed and a test data set to illustrate the protection methodology.  Alternatively, secure enclaves~\cite{secure_enclaves} or multi-party computations~\cite{mpc} can be adopted to obscure the user's ability to view certain aspects of the data without data editing or suppression.

With Federal agencies, some programs require data users to write research proposals and make those proposals public. For example, the All of Us Research Program~\cite{all2019all} publishes research proposals for a public audience. For some other programs, there is a process, set by the FDA. The user's research questions in the process affect the user’s privacy budget. Review committees often determine which questions are allowed. Clearly, there is a distinction between a committee review and a larger population which has greater visibility into what is happening. Since data subjects are compelled to contribute data, they are entitled to know how their data are being used. This level of transparency may imbue trust from data subjects. However, there might be some boundary conditions that make this level of transparency an impracticable option.

\subsubsection{Incentive Structures}
To encourage data sharing and taking on risks, incentives such as assigning more computing power or monetary values to those who share are very effective. Incentive structures will be key for stakeholder engagement. For data subjects, incentives affect the participation rate in the data collection process. For data stewards, incentives affect the sharing rate of collected data. For example, the Defense Advanced Research Projects Agency~(DARPA) collects large amounts of data during research programs~\cite{RP17}, however, this data is rarely made available to other institutes or to the public. In the case where data are changed or updated in one place after being shared, incentives are needed to propagate the change across data sets.  An example can be a national research cloud; where researchers upload research data to be shared with others.  The ownership of privacy risk is unclear in a distributed cloud system and the privacy risk assessments need to take into account data subjects and affected groups.

 The impacts of incentives on a stakeholder's utility function and the corresponding optimal settings could be examined and obtained using game theory or mechanism design approaches under different types of assumptions. However, in practice, it might be challenging to elicit the dimension, constraints, and considerations from stakeholders to inform the optimization for the utility functions. In addition, there might be contention between academic and Federal data stewards. Currently, there is a push in the Department of Defense~(DOD) to incentivize military members to share healthcare data. However, DOD needs to review datasets for potential adverse events to the nation, which might be different from the way academics may review such data in terms of risk assessment.


\subsection{Equity} \label{subsec:equity}
Existing mechanisms for disclosure limitations for data sharing can have an outsized impact on minority populations. For example, suppression methods may completely hide rare medical diagnoses; differential privacy can result in analyses that in percentage terms are substantially less accurate for small populations~\cite{10.1561/0400000042}. The use of these disclosure limitation mechanisms can potentially cause representative inequities, and the minority populations are likely to be under-represented in the data being used in scientific research. Representation inequities can lead to systemic biases in the decisions based on the data and can further lead to greater errors for under-represented groups. This may impact the ability to direct resources to these under-represented groups.

At the same time, re-identification risks are not equal among all persons in a dataset. For example, a large proportion, if not all, of the people at high risk are usually those with small group sizes for demographic or health-related characteristics. Consequently, the burden of re-identification risk disproportionately falls on traditionally disadvantaged groups, including people with rarer health conditions and members of minority racial or ethnic groups~\cite{Simon2019}. These individuals often face greater risk of harm from re-identification (\textit{e.g.}, potential discrimination based on expectation of future healthcare costs). 

Some of the workshop participants felt the conflicting objectives of reducing the re-identification risk and increasing representation in the data for historically disadvantaged groups raised significant challenges and was under-explored in existing research in data privacy in general and disclosure limitation specifically. Two sessions discussed these issues; the participants of the workshop had the following findings, mentioned below:

\paragraph*{Problems of poor quality results for minority populations caused by data inequities is not limited to privacy protection methods.}
The outcome inequities for minority populations are not always the artifacts of privacy protection or disclosure control methods; there are also inherent inequities and the impact of overall system design (such as using
VA healthcare data as a surrogate for the general population). It is important to distinguish between the different sources of data inequities. 
One example brought up in the discussion during the workshop was county-level COVID-19 statistics, in which minorities are likely to be under-represented. This is because the data only include test-confirmed COVID-19 cases, and minority communities may be likely to have less testing facilities. Further, minorities may be more reluctant to take a test, due to reasons such as lacking insurance and a history of being unfairly treated by health professionals. The under-representation could lead to an under-estimation of the prevalence of the disease in minority communities, which results in the under-representation of the minority groups in the data that will be used in health research related to COVID-19.

Oversampling is a mechanism that has been used for addressing the latter. However, when one thinks of subpopulations that are not necessarily defined in terms of these common design variables, that is when a subpopulation's definition involves other variables which are not design variables, then these subpopulations cannot be oversampled  (because the values of non-design variables are not known until the data is collected); in these particular cases, representation and disclosure risk is really a problem. Yet another related issue is: if the data is high dimensional can we even identify all these subpopulations/groups that might be significantly different from the others in terms of certain sensitive variables and thus require adjustments in disclosure limitation protection in order to offer the same level of protection to these groups? For example, certain health conditions are rare among certain subpopulations of individuals, but not so rare in others. In theory, one can define a very large number of subpopulations defined by the cross-classification of the levels of the variables in the data set. Machine learning techniques, such as association rule mining can be helpful for that~\cite{mult_top}.  However, more research is necessary.

There are also model-based techniques, such as small-area estimation~\cite{SmallAreaEstimation} that can be used for data with finer geographic detail in published statistics and for various subpopulations. Traditional demographic sample surveys designed for national estimates do not provide large enough samples to produce reliable direct estimates for small areas such as counties and even most of the states. Small area estimation techniques can help improve the precision of analysis on small subpopulations. This raises the research question of whether such techniques can be used to reduce the impact of privacy protection mechanisms? Research on developing small area estimation models for analyzing small subpopulations on data altered by privacy disclosure control methods needs to be done. A second question, that again is closely related to non-privacy issues, is convincing stakeholders of the validity of results from such methods.
    
\paragraph*{Multiple Methods for Data Access.}
It may be difficult to release a dataset with unlimited/uncontrolled access to adequately address privacy concerns while still providing for analyses on small population subgroups with sufficient fidelity. This problem also arises when analysis requires linking records across multiple datasets. For example, the COVID-19 Case Surveillance Public Use Data from the Centers for Disease Control and Prevention (CDC)~\cite{Garcia2020} may contain information about pre-existing conditions and co-morbidity that are unique to an individual and may only be used anecdotally and not for statistical analysis purposes. But this information may become useful for statistical analysis when the CDC data are linked with other data sources. Privacy protection methods can make this particularly difficult.
The burden of privacy protection falls on the data access control mechanisms, when applying disclosure control methods to the data will cause issues in conducting analysis or linkage related to small population subgroups.

Controlled/limited data access approaches, such as in Research Data Centers (RDC), and synthetic data combined with validation servers discussed in the Day 1 talks~(Table~\ref{tab:sch1}), could potentially provide sufficient privacy protection allowing for analysis of small population subgroups. For example, the National Center for Health Statistics~(NCHS) operates RDCs to allow researchers access to restricted-use data. The RDC is responsible for protecting the confidentiality of survey respondents, study subjects, or institutions while providing access to the restricted-use data for statistical purposes. For access to the restricted-use data, researchers must submit a research proposal outlining the need for restricted-use data. The proposal provides a framework for NCHS to identify potential disclosure risks and how the data will be used.

Synthetic data could be used for computing simple statistics but may not be reliable for producing estimates for complex statistical models. Validation servers allow the execution of statistical programs developed and debugged on synthetic data to run using confidential data with noise added to estimates to preserve privacy~\cite{9138552}.

Applying the controlled/limited data access approach raises four research questions:
\begin{enumerate}
    \item How to determine whether such alternatives need to be provided. This involves determining whether there will be issues with analysis of small population subgroups using the data.
    \item How to evaluate the potential disclosure risk based on the profile of the researcher that requests the data and the how the data will be used as specified in the research proposal.
    \item How to generate statistically valid estimates for small group analysis with robust measurable privacy protection using synthetic data and validation servers.
    \item How to embed equity as a key factor into the design of the synthetic data and validation server system.  For example, if a validation server has a limited ``privacy budget'', a portion of that should be reserved for validating analyses of minority populations where adequately accurate analyses cannot be performed through other means.
\end{enumerate}

\paragraph*{Communicating Best Practices.} It is important that users of data understand the potential for inaccurate results on small population subgroups. There are likely to be a suite of statistical models and analysis tools available to perform analyses on small population subgroups, such as random effects models, the Fay-Herriot model~\cite{BENAVENT2016372}, Bayesian pseudo-likelihood models, and synthetic data combined with validation servers. Different tools and models provide different precision in model estimates for different analysis.  Therefore, it is critical to understanding which tool to use for what purpose. This is an issue both for the data controller, who must make sure appropriate capabilities are available for analysis of small subgroups/minority populations, and data users, who must be able to choose appropriate tools for the analyses they are performing.

This is a part of a broader issue of determining credibility of analyses, particularly with data protected through privacy technologies. The equity of such analyses, and the decisions made based on such analyses, needs to be considered as part of such credibility issues. In the Artificial Intelligence community there are many research efforts underway to understand and characterize various aspects of fairness and develop tools for assessing and improving the fairness of the AI systems. For example, AI Fairness 360 (AIF360)~\cite{8843908} is an open-source Python toolkit for algorithmic fairness, which includes a comprehensive set of fairness metrics for datasets and models, explanations for these metrics, and algorithms to mitigate bias in datasets and models. Looking into these tools may provide insight into what is the best practice for data analysis for small population subgroups, when privacy disclosure control techniques are applied. 

The workshop arrived at these ideas for tackling the issues with equity and credibility associated with the data analysis tools and models:
\begin{enumerate}
    \item Develop a set of use cases in which the different privacy disclosure control methods are applied to the data, and show the impact on the accuracy of the analysis results using the data (e.g., ROC curves) for different population subgroups.  Given various methods, the impact on data equity for each of the methods should be mentioned. Even though in many cases, the impact can be obvious, there are also cases in which the impact on data equity can be complex. Use cases that represent the scenarios in which the data analysis approach affects the data equity in a complex way can be helpful for  data controllers.   
    \item Develop a set of use cases in which different analysis tools and models are used on the data and show the impact on the accuracy of analysis results for population subgroups of different sizes. The workshop participants suggested particular examples of use cases, such as the use case of the COVID-19 Case Surveillance Public Use Data from the CDC, and the data collected by the DARPA.
    \item Develop tools for characterizing types of uses that could cause issues: for example, data analysis at highly granular geographic levels, analysis of high dimensional data, and analysis of minority population subgroups.
    \item Training sessions at data users' conferences and other venues. Developing such a tutorial could be cross-agency and cross-data-set.
    \item Work with journal editors to make sure they are aware of potential issues and understand how to spot likely misuse of tools that would lead to questionable results.
    \item Develop model diagnostic and model comparison tools for small population subgroup analysis.
\end{enumerate}

\paragraph*{Variable Fidelity Methodologies.} The goal of privacy disclosure control methods is essentially to find an optimal trade-off between disclosure risk and fidelity of the data. Individuals from small population subgroups are inherently more likely to be identified or to have their sensitive information disclosed.  If disclosure control methods convert them to the same granular level or fidelity level as individuals from the larger population groups, there are two possible outcomes. One, if the granular level is controlled at a level to ensure the minority groups have sufficiently low risk, the data from the majority groups would lose fidelity unnecessarily. Two, if the granular level is controlled at a level that the risk-utility trade-off is optimal for the majority groups, it is likely that the minority groups would not be sufficiently protected. A potential way to address this issue is to apply disclosure control methods that allow for different levels of fidelity for different groups. For example, the local-recoding~\cite{10.1145/1150402.1150499} based anonymization method searches for solutions in a space in which the same value of a attribute can be generalized to different levels for different individuals. For example, if two people are both 40 years old, but one of them is White and the other American Indian or Alaska Native~(AIAN), the local-recoding method can generalize the age of the AIAN person to a large age group (\textit{e.g.}, 30-50) while keep the age of the White person as is. Applying the variable fidelity methodologies raises the following research questions:

\begin{enumerate}
    \item Develop privacy disclosure control methods that allow varying levels of fidelity for individuals from different population subgroups that optimally trade-off between privacy risk, equity and data fidelity, or data utility;
    \item  Develop materials for communicating the need for, and benefit from, such varied privacy level approaches.
\end{enumerate}
A related problem is when data are inherently non-representative for statistical analysis but may still be used to address equity issues. An example was given for CDC COVID-19 data about pre-existing conditions and co-morbidities that are unique to an individual but may be used anecdotally to help address equity issues.

\paragraph*{Understanding Stakeholders.}
It is important to recognize that stakeholders (both the data subjects and those impacted by the decisions made on the basis of the data) from different sub-populations may have different concerns; the challenge is to understand the concerns of each group. Even defining populations can be difficult~\cite{COBB2016595}, for example self-identification as AIAN vs. official tribal affiliations. Understanding how data will be used and who will be impacted is critical to determining whether technologies used are appropriate.

Ideas include standing advisory councils, publishing notices of data sharing practices, and putting out ``experimental'' versions of data to see the response within the stakeholder community. One suggestion was to concentrate not on the negative impact of privacy protection, but frame problems as ``Here is where we stand, here are areas we want to improve'' or ``We currently do not have the ability to do analyses for group, can we get something?''

\paragraph*{Ethics in Data Collection and Use}
Equity issues raise special ethical concerns, and these need to be understood and addressed. Ethics issues, and particularly equity issues, should be incorporated into any training required for data access. Issues raised include:
\begin{enumerate}
    \item What are the costs of gathering as well as using the data, and how does this vary among subgroups?
    \item Where does accountability lie when different groups may be impacted in different ways by privacy technology?
    \item How do we deal with information asymmetry between those making decisions with respect to a particular subgroup vs. those looking at larger groups?
\end{enumerate}

The participants also brought up that these issues may not be limited to person-related data. For example, the talk on the salary equity study done by the Boston Women's Workforce Network showed improvement in salary equity, but the design of the privacy protection mechanism may preclude ``drilling down'' to determine whether that improvement is confined to particular business sectors or salary ranges.

\subsection{Technology Communication} \label{subsec:tech}
Guidance for privacy technology – there is a need for proper guidelines (in addition to standards, if possible) that can provide guidance to agencies/stakeholders with regards to which privacy solutions best fit each agency's or inter-agency's needs. There is also a need to establish ways to foster better understanding of the pros and cons of different privacy technologies for different use cases/scenarios and as contexts change. Best practices or standards need to be explored.

There is a need to develop a clear understanding of adversarial threats to privacy within each agency or use case so as to provide measured solutions that balance privacy and utility based on well-understood risks.

Privacy in the government sector is very different than in industries; there is a need to foster understanding that privacy in government is fundamentally different than in the context of industry. Users have opportunities to ``decide'' whether to share data with companies (\textit{e.g.}, Facebook, etc.), but data collection by governmental organizations is mandated in many cases because of citizenry. So there is a higher standard of trust and protection expected from governmental units as there is a broader impact on society for the loss of privacy protection; in general the stakes can be higher in the governmental sector.

Success stories demonstrate modern privacy technologies can be leveraged – there was a feeling that privacy practice (or practical deployment) lags about 20 years behind privacy theory (at least in some cases); but we have success stories (\textit{e.g.}, Census/IRS) that show that with concerted efforts current research outputs can be transitioned to practice successfully.

\subsection{Methods} \label{subsec:methods}

\subsubsection{Existing work}

Existing work on privacy-preserving data sharing falls largely within two areas. The first area concerns privacy of the output, addressing sensitive information leakage by computational outputs about confidential data inputs. Techniques like differential privacy~\cite{dwork2014algorithmic} and statistical disclosure limitation~\cite{matthews2011data} fall within this area. Alternatively, the second area concerns privacy of the computation, addressing the security risks of actually performing the computation on one or multiple machines that may or may not be trusted with some or all of the confidential data. Techniques like fully homomorphic encryption~\cite{martins2017survey} and secure multiparty computation~\cite{du2001secure} fall within this area. These two areas are orthogonal to one another; techniques that address one cannot solely be used to address the other, and vice versa.


Protecting privacy of the input requires balancing a delicate trade-off between disclosure risk and utility of any disclosed results. In order to quantitatively assess this trade-off, assumptions must be made about possible data generating scenarios and potential adversarial knowledge~\cite{kifer2011no}. Once these assumptions are made, data curators need to choose measures of disclosure risk and data utility to quantitatively assess the trade-off. Many privacy frameworks quantify this trade-off parametrically (for example, the ``$\varepsilon$'' in $\varepsilon$-differential privacy or the ``k'' in $k$-anonymity). Because of the diversity of stakeholder needs for any data sharing practice, no single set of assumptions, metrics, or framework choices can broadly resolve these challenges. This makes it particularly difficult to set standards and best practices for private data sharing, even when reasonable assumptions allow for the trade-off to be parametrized~\cite{lee2011much}. Problems like these highlight the benefits of tiered access to confidential data. For example, $\varepsilon$-differentially private synthetic data could be released publicly with strong privacy protection (\textit{i.e.}, smaller $\varepsilon$), and with weaker privacy protection (\textit{i.e.}, larger $\varepsilon$) to authenticated researchers generically or to authenticated researchers for a specific use who need to work directly on confidential data in an enclave or other secure environment.


Similar issues arise for implementing privacy of the computation; preserving the privacy of the computation incurs a new trade-off between disclosure risk and computing time and resources. The effectiveness of private computing techniques similarly depends on assumptions about the trustworthiness of different computing resources and the channels used to communicate. Often times, these assumptions are hard to validate in practice; for example, different cloud computing platforms may have different privacy policies regarding data storage and use, and the actual risks may be less transparent.

\subsubsection{Agencies’ needs and open questions}

Through the discussions during the workshop breakout session on methods, a number of key methodological needs have been identified for agencies hoping to implement private data sharing practices. These map broadly onto open questions in new areas of research that we believe best help agencies meet these needs. Generally speaking, the goal is to use methodological research to close the gap between academic research on privacy-preserving data sharing and the existing needs and wants of data sharing agencies.

\begin{enumerate}
    \item \textbf{Improved matching of existing methods to specific utility needs}: Different stakeholders will inevitably negotiate and agree on the trade-offs between privacy, utility, and computation described above. Many privacy frameworks do not offer easily interpretable privacy and/or utility guarantees, making it difficult for agencies to identify methods that best fit their use cases. As a result, the negotiation process often happens indirectly through empirical analyses. For example, when the US Census Bureau decided to adopt differential privacy for its 2020 decennial releases, they held multiple workshops where data users would perform their analyses on prototype private data and provide feedback on how well the prototype data suited their use case~\cite{kirkendall20202020}. This iterative process allowed the US Census Bureau to update their methodology to better meet these use cases. However, it also highlights a need for better communication with data curators about their data sharing and usage needs, so that they can be best translated into a privacy-preserving end product. 
    
    Organizational processes like these also help agencies define risk and utility priorities more directly. Not all agencies will need or even want to protect against the same kind of attacks, and not all agencies will have the same practical utility needs. These points are often overlooked in the privacy methods literature. For example, with differential privacy, the majority of the existing literature compares methods using expected errors induced by privacy alone; however, this does not tell agencies how these choices affect statistical inference and decision-making using these data, nor does it say how to propagate uncertainty through these inferential tasks. 
    
    One potential avenue of addressing these needs more efficiently is to consider fitness-for-use models, which solve the dual problem of maximizing a measure of privacy preservation while maintaining a fixed amount of utility with respect to some metric (see \cite{xiao2020optimizing} for an example on private linear queries similar to the US Census Bureau use case). This allows data curators to specify utility constraints as inputs and use these to infer private release strategies as outputs. While methods like these can more directly match user needs with private methods, they can also pose new challenges for research. When privacy parameters are tuned based on functions of the confidential data, the resulting privacy parameter can implicitly depend on the confidential data, yielding potential disclosure risks. This has been a major barrier to hyper-parameter tuning using techniques like differential privacy.
    
    \item \textbf{New techniques for easing implementation burdens of privacy-preserving data releases}: Techniques for privatizing computations and outputs need not always be compatible; in fact, implementing both independently can lead to scenarios that technically satisfy the assumptions of frameworks such as secure multiparty computation and/or differential privacy, but may otherwise defeat the spirit of both~\cite{clifton2013challenges}. Furthermore, techniques for privatizing the output of statistical queries often rely on highly sanitized academic settings in which implementing the ideal algorithm may not be cryptographically secure in practice. Academic research in these areas tends to target publications in more theoretical outlets, for which it is commonplace to ignore implementation barriers when deriving results. For example, differentially private releases can be susceptible to attacks when continuous arithmetic is assumed, but in practice is approximated by double-precision arithmetic~\cite{mironov2012significance}. 
    
    These issues highlight a gap between theory and practice when it comes to development of privacy-preserving methods. Closing this gap is simultaneously a methods problem and a systems problem; policy decisions, like those involved in establishing tiered access systems, may require new ways of integrating policy into mathematical methods. 
    
    \item \textbf{Methods for more realistic measurement of disclosure risk}: Characterizing disclosure risk in practice remains quite challenging. For example, differential privacy can provide relative robustness to record linkage attacks on databases. However, many practical aspects of the linkage attack are difficult to quantify, including but not limited to the computational efficacy of the attack, assumptions about precise attacker knowledge, logistical or policy-based constraints, and verification of attack success. Similarly, in cryptography, there is a nebulous relationship between adversary assumptions (honest, semi-honest, or malicious) and the practicality of executing a cryptographic attack.
    
    More generically, modeling disclosure risk from privacy often requires comparing multiple metrics contextually. Relative privacy guarantees do not allow for ``apples-to-apples'' comparisons with individual disclosure risk~\cite{mcclure2012differential} or feasible, verifiable re-identifications~\cite{domingo2021limits}. Moreover, agencies are likely to borrow techniques from multiple frameworks above, many of which are closely associated with one particular risk measure. Agencies need a more holistic view of disclosure risk in order to establish better data sharing policies.
    
    \item \textbf{Systems-level effects of multi-tiered access}: Multi-tiered access can ease many logistical burdens associated with data sharing by matching users to different access levels based on their own trustworthiness and the data needed to perform their tasks. However, modeling disclosure risk in a multi-tiered access situation becomes exponentially more challenging. In terms of both the computation and the output, assumptions must be made not only about each involved party, but how said parties interact with each other. For example, the US Census Bureau is currently investigating private synthetic data with validation servers for public releases of the American Community Survey~\cite{freimanformal}. However, the Census Bureau already has well-established research data centers where trusted users can work directly on confidential data. These users, either communicating directly through sharing data or indirectly through sharing results, can potentially pose new challenges for privacy-preserving risk analysis. 
    
    Furthermore, multi-tiered access creates important policy problems that need to be addressed in practice. Protecting privacy in practice usually requires finite resources of some kind; these could be query limits in a cryptographic setting or a privacy budget in a differential privacy setting. Regardless, no active privacy-preserving system can be infinitely reused and still maintain the privacy  guarantees originally specified. Treating these as finite resources require systems-level analysis to set both technical and social policies around how best to allocate these resources. Ideally, we want to avoid scenarios in which systems must be taken down because such finite resources are depleted. 
    
\end{enumerate}

\section{Actionable Items}\label{sec:recommendations}
While the workshop did not reach any specific conclusions or make specific recommendations, the authors of this report feel that progress is needed on several fronts.

\paragraph*{Problem Understanding} is a key challenge.  There are a variety of constraints that limit data sharing, and a wide variety of requirements for data.  This leads to a myriad of solutions, some based on existing technology, but others potentially demanding research and engineering advances. Several activities that could be supported to ensure advances include:

\begin{description}
  \item[$-$ Testbeds (inclusive of Datasets):]  Having a common repository with a variety of data sharing use cases may stimulate research. A use case should include descriptions of data, the concerns/constraints on releasing that data, and discussion of how the data would be used if shared. Description of data should cover the topics such as data schema, data collection protocols and data generating processes. Wherever feasible, a miniaturized real or synthetic sample dataset along with specific examples of how data has been used and agreements required for that sharing would be valuable.
  
  \item[$-$ Areas of Focus:] Based upon the discussions held during and after the workshop,  areas of focus suggestions are provided to the community. Before discussing the specifics, these areas are categorized into foundational and applied. Research advancement in both areas would benefit each other in a closed-loop manner. Applied research suggestions are discussed below: 
  \begin{itemize}
      \item Application-oriented data release: Releasing data in an application-independent manner could come at a cost of a low utility and/or high disclosure risk. Therefore, more research is necessary into task-oriented data releases that could alleviate low privacy-utility concerns. This would require the involvement of subject area experts (e.g., public health experts or economists) to achieve a better understanding of the relevance and importance of different applications. Privacy experts would need to develop bespoke mechanisms that are specifically focused on boosting utility for a given application. Collaboration with qualified statisticians and/or data scientists would be necessary when the goal is to release government survey data, for example, there is an increasing need in defining methods that would satisfy rigorous privacy criteria and that can be applied to complex survey data.  Such applications could be direct collaborations on a specific data release or could use testbed or contest-style frameworks to elicit multiple approaches to addressing a particular application.
      \item Modular interfaces for privacy-preserving data analysis: The translation time from basic research to real-world application for privacy-preserving methods can be quite long. This time until translation can be significantly shortened by developing libraries and frameworks that set standards for the usage of privacy-preserving algorithms. Similar initiatives by the Internet Engineering Task Force (IETF)~\cite{bradner1999open} have led to Request For Comments (RFC) that accelerates the adoption of innovative work in cryptography and other internet technologies. 
      \item To facilitate greater acceptance of synthetic data by data users and wider adoption of synthetic approaches by privacy-preserving practitioners,  it would be desirable to have a sound methodology of verification servers that can be applied to the real-world scenarios of government data release in practice.
  \end{itemize}
  
  At the same time, several opportunities for foundational research are discussed below:
  \begin{itemize}
      \item Algorithms for quality assessment: Assessing data quality is a crucial step in any application of data-driven decision making. Typically, evaluation is performed under the assumption that data is accessible for evaluation. However, in privacy-preserving data release, this assumption may not be true. Therefore, analyzing the properties of data and data distributions when there is no access to raw data introduces a new set of research challenges. Addressing such challenges could allow practitioners to evaluate databases for data quality and allow them to ask queries for which sufficient quality can be retained while preserving privacy.
      \item Further research is needed to develop metrics that quantify factors associated with the privacy of data such as information leakage, identifiability, and usability in a principled and mathematical framework.
      \item Access management tools for data privacy: Sharing data always comes with a certain utility-privacy tradeoff. Research in the development of metrics and tools of risk-utility assessments is needed to make it easier for various stakeholders involved in the data sharing process to understand and balance the risk associated with data release and the benefits of data release to society.
      \item Distributed data query pools: An alternative to releasing datasets is to develop a secure outsourced computation infrastructure that instead of sharing raw data performs computation on data and returns the answer. Several recent advancements in secure multiparty computation and trusted execution environments can act as a building block. Still, more research is needed to develop algorithms and systems that privately and securely answer interactive queries from data analysts that are acceptable to the framework of government data sharing.
      \item Taxonomy of private data: Data privacy is a multifaceted challenge where individuals, government, organizations, and data analysts have different expectations and challenges. Therefore, research in the taxonomy of privacy and usability concerns associated with data can improve the adoption of private data release techniques. 
  \end{itemize}

  \item[$-$ Workforce \& Incentives Problem:] People are needed who can understand the problems and applications of data sharing and associated technologies. We note that this is not merely about technologists and educating technologists, it is also about educating people from a variety of disciplines and backgrounds so that they understand the privacy issues, the data-sharing issues, the technologies they use and its direct or indirect impact. In particular, engaging the stakeholders by educating people from a variety of disciplines so that they understand the privacy issues, the data-sharing issues, and the data driven technologies they use.
  \item[$-$ Challenges:] It is important to have experts that understand the potential harms associated with sharing a particular type of data and experts that understand the utility of that data. Collaboration between such experts would enable understanding of the particular cases, what the data are and the types of data that people have not thought of, which can help in addressing the problems that people are trying to solve. We consider the example of privacy-preserving machine learning in this context. If one’s goal is to use machine learning models, data users may not get access due to restrictions on the usage of the machine learning model or restrictions due to lack of access over the data the machine learning model was trained on. Therefore, it is important to know, what is the goal? What does it mean in terms of data access control policies to do that? And likewise, what are the risks? Can we quantify those harms that can happen? Can we better understand the perceptions? This will help us to understand the distinction between the perceived risks and the actual potential for harm.
  
  \item[$-$ Policy Research:] One of the first things that came out of the workshop is there is a need for policy research that is outside the scope of workshop. How do we understand the harms at individual and community levels, stepping beyond the disclosure risks? What’s the harm from disclosure and who bears the brunt? How does it impact them? What are perceived harms? How does that compare to the real harm? Are there benefits to risk? We do not know the best way to communicate such issues. As we know that with the Census Bureau's work in adopting differential privacy for releasing their data, one of their issues has been communicating what this means, both in terms of how does it meet their mandate for protecting privacy, but also how does it impact the fissures of data?
  
  There is a big concern that differential privacy is going to decrease the data's utility for practical purposes. This concern is juxtaposed with another set of concerns that differential privacy would not be sufficient for providing privacy at a group or community level. Researchers' use of such technologies mostly has been inside the lab and not stress tested with real world constraints. This leads to a communication gap, which begs numerous questions. In particular, how can we communicate the risks, the benefits, and the technologies and finally, how do we turn this into policy? These questions aim to bring research questions to the surface that need to be addressed by the technology and policy community by working together and going beyond development of policy and technology in isolation. How can we move from understanding some of these phenomena to developing actionable policy? This is certainly something that could be a low-risk high-reward investigation. If the challenges and the problems, limitations, and constraints can be described and then used to build a community around particular use cases or classes of use cases, a means to address those issues could be developed.
  
  \item[$-$ Competitions:] Competition can stimulate advancement in technologies by bringing together disparate scientists and engineers. However, when sponsoring these endeavors, one must ask what is the goal? If the goal is to advance a solution to a very particular problem and the challenge or competition is that problem, then can one look for what is the state-of-the art out there today to address that problem? Or does it require development of completely new technologies? One way that often works best is by building a community by organizing a series of conferences and workshops where people can participate, discuss the problems, and discuss how to further work on these as a community. This can be thought of in terms of community building by considering what infrastructure is needed to build around that challenge to ensure that the community is not just solving toy problems with limited generalizability or that there are opportunities to transition into practice.
  
\end{description}


\section{Acknowledgement} The authors thank the National Science Foundation under award nos. 2129895, 2129895, 2129970, 2115149 and 2129909, the National Institute of Standards and Technologies, and the White House Office of Science and Technology Policy for their support.

\bibliographystyle{plain}
\bibliography{ref}

\begin{thebibliography}{10}

\bibitem{UCIrepository}
A.~Asuncion and D.J. Newman.
\newblock {UCI} machine learning repository.
\newblock http://www.ics.uci.edu/$\sim$mlearn/{MLR}epository.html, 2007.

\bibitem{8843908}
R.~K.~E. Bellamy, K.~Dey, M.~Hind, S.~C. Hoffman, S.~Houde, K.~Kannan,
  P.~Lohia, J.~Martino, S.~Mehta, A.~Mojsilović, S.~Nagar, K.~Natesan
  Ramamurthy, J.~Richards, D.~Saha, P.~Sattigeri, M.~Singh, K.~R. Varshney, and
  Y.~Zhang.
\newblock Ai fairness 360: An extensible toolkit for detecting and mitigating
  algorithmic bias.
\newblock {\em IBM Journal of Research and Development}, 63(4/5):4:1--4:15,
  2019.

\bibitem{BENAVENT2016372}
Roberto Benavent and Domingo Morales.
\newblock Multivariate fay--herriot models for small area estimation.
\newblock {\em Computational Statistics \& Data Analysis}, 94:372--390, 2016.

\bibitem{bradner1999open}
Scott Bradner, J~Hamerly, K~McKusick, T~O’REILLY, T~Pacquin, B~Perens,
  E~Raymond, R~Stallman, M~Tiemann, L~Torvalds, et~al.
\newblock Open sources: Voices from the open source revolution, 1999.

\bibitem{clifton2013challenges}
Chris Clifton and Balamurugan Anandan.
\newblock Challenges and opportunities for security with differential privacy.
\newblock In {\em International Conference on Information Systems Security},
  pages 1--13. Springer, 2013.

\bibitem{COBB2016595}
Ryon~J Cobb, Courtney~S Thomas, Whitney~N {Laster Pirtle}, and William~A
  Darity.
\newblock {Self-identified race, socially assigned skin tone, and adult
  physiological dysregulation: Assessing multiple dimensions of “race” in
  health disparities research}.
\newblock {\em SSM - Population Health}, 2:595--602, 2016.

\bibitem{secure_enclaves}
Victor Costan, Ilia Lebedev, Srinivas Devadas, et~al.
\newblock {\em Secure processors part I: background, taxonomy for secure
  enclaves and Intel SGX architecture}.
\newblock Now Foundations and Trends, 2017.

\bibitem{domingo2021limits}
Josep Domingo-Ferrer, David S{\'a}nchez, and Alberto Blanco-Justicia.
\newblock The limits of differential privacy (and its misuse in data release
  and machine learning).
\newblock {\em Communications of the ACM}, 64(7):33--35, 2021.

\bibitem{du2001secure}
Wenliang Du and Mikhail~J Atallah.
\newblock Secure multi-party computation problems and their applications: a
  review and open problems.
\newblock In {\em Proceedings of the 2001 workshop on New security paradigms},
  pages 13--22, 2001.

\bibitem{10.1561/0400000042}
Cynthia Dwork and Aaron Roth.
\newblock The algorithmic foundations of differential privacy.
\newblock {\em Found. Trends Theor. Comput. Sci.}, 9(3–4):211–407, August
  2014.

\bibitem{dwork2014algorithmic}
Cynthia Dwork, Aaron Roth, et~al.
\newblock The algorithmic foundations of differential privacy.
\newblock {\em Found. Trends Theor. Comput. Sci.}, 9(3-4):211--407, 2014.

\bibitem{9138552}
Khaled El~Emam.
\newblock Seven ways to evaluate the utility of synthetic data.
\newblock {\em IEEE Security Privacy}, 18(4):56--59, 2020.

\bibitem{FCCPP}
{FCC}.
\newblock Fcc speed test app privacy policy.
\newblock
  https://www.fcc.gov/general/mobile-broadband-performance-application-privacy-notice-and-terms-use,
  2021.

\bibitem{Patent1790}
{First Congress}.
\newblock an act to promote the progress of useful arts.
\newblock
  https://www.loc.gov/law/help/statutes-at-large/1st-congress/session-2/c1s2ch7.pdf?loclr=bloglaw,
  April 10 1790.

\bibitem{freimanformal}
Michael~H Freiman, Rolando~A Rodr{\'\i}guez, Jerome~P Reiter, and Amy Lauger.
\newblock Formal privacy and synthetic data for the american community survey.
\newblock {\em NCES report}, 2018.

\bibitem{Garcia2020}
Macarena Garcia, Nikolay Lipskiy, James Tyson, Roniqua Watkins, E~Stein Esser,
  and Teresa Kinley.
\newblock {Centers for Disease Control and Prevention 2019 novel coronavirus
  disease (COVID-19) information management: addressing national health-care
  and public health needs for standardized data definitions and codified
  vocabulary for data exchange}.
\newblock {\em Journal of the American Medical Informatics Association},
  27(9):1476--1487, sep 2020.

\bibitem{mpc}
Oded Goldreich.
\newblock Secure multi-party computation.
\newblock {\em Manuscript. Preliminary version}, 78, 1998.

\bibitem{kifer2011no}
Daniel Kifer and Ashwin Machanavajjhala.
\newblock No free lunch in data privacy.
\newblock In {\em Proceedings of the 2011 ACM SIGMOD International Conference
  on Management of data}, pages 193--204, 2011.

\bibitem{kirkendall20202020}
Nancy~J Kirkendall, Constance~F Citro, and Daniel~L Cork.
\newblock {\em 2020 Census Data Products: Data Needs and Privacy
  Considerations: Proceedings of a Workshop}.
\newblock National Academies Press (US), 2020.

\bibitem{law3024law}
Public Law.
\newblock Law 114--255—dec. 13, 2016 130 stat. 1033, 21st century cures act,
  12/2016.

\bibitem{evidence}
Public Law.
\newblock Public law 115–435—jan. 14, 2019 132 stat. 5529, the foundations
  for evidence-based policymaking act of 2018 ("evidence act"), 01/2019.

\bibitem{lee2011much}
Jaewoo Lee and Chris Clifton.
\newblock How much is enough? choosing $\varepsilon$ for differential privacy.
\newblock In {\em International Conference on Information Security}, pages
  325--340. Springer, 2011.

\bibitem{levenstein2020role}
Margaret Levenstein.
\newblock Role of federal statistical research data centers in promoting
  transparency and reproducibility in federal statistics.
\newblock In {\em Joint Statistical Meetings of the American Statistical
  Association}, August 2020.

\bibitem{martins2017survey}
Paulo Martins, Leonel Sousa, and Artur Mariano.
\newblock A survey on fully homomorphic encryption: An engineering perspective.
\newblock {\em ACM Computing Surveys (CSUR)}, 50(6):1--33, 2017.

\bibitem{matthews2011data}
Gregory~J Matthews and Ofer Harel.
\newblock Data confidentiality: A review of methods for statistical disclosure
  limitation and methods for assessing privacy.
\newblock {\em Statistics Surveys}, 5:1--29, 2011.

\bibitem{mcclure2012differential}
David McClure and Jerome~P Reiter.
\newblock Differential privacy and statistical disclosure risk measures: An
  investigation with binary synthetic data.
\newblock {\em Trans. Data Priv.}, 5(3):535--552, 2012.

\bibitem{mironov2012significance}
Ilya Mironov.
\newblock On significance of the least significant bits for differential
  privacy.
\newblock In {\em Proceedings of the 2012 ACM conference on Computer and
  communications security}, pages 650--661, 2012.

\bibitem{all2019all}
All of~Us~Research Program~Investigators.
\newblock The “all of us” research program.
\newblock {\em New England Journal of Medicine}, 381(7):668--676, 2019.

\bibitem{mult_top}
Anna Oganian, Ionut Iacob, and Goran Lesaja.
\newblock {Multivariate Top-Coding for Statistical Disclosure Limitation}.
\newblock In Domingo-Ferrer J and Muralidhar K., editors, {\em Privacy in
  Statistical Databases, Lecture Notes in Computer Science}, volume 12276,
  pages 136--148. Springer-Verlag, 2020.

\bibitem{SmallAreaEstimation}
J.N.K. Rao.
\newblock {\em Small Area Estimation}.
\newblock John Wiley and Sons, Inc., Hoboken, New Jersey, 2003.

\bibitem{RP17}
{Regli, William C. and Pierce, Brian}.
\newblock Darpa and data: A portfoli overview.
\newblock https://www.nitrd.gov/nitrdgroups/images/3/31/DARPA-and-DATA.pdf,
  2017.

\bibitem{ridgeway2020crisis}
Diane Ridgeway, Christine Task, Gary Howarth, and David Van~Ballegooijen.
\newblock Crisis collaborations: Challenges for safe data sharing with
  differential privacy.
\newblock {\em NIST (PSCR 2020)}, 2020.

\bibitem{NISTDPSyn}
Diane Ridgeway, Mary~F. Theofanos, Terese~W. Manley, and Christine Task.
\newblock Challenge design and lessons learned from the 2018 differential
  privacy challenges.
\newblock Technical Report NIST Technical Note 2151, National Institute of
  Standards and Technology, April 2021.
\newblock \url{https://doi.org/10.6028/NIST.TN.2151}.

\bibitem{Simon2019}
Gregory~E Simon, Susan~M Shortreed, R~Yates Coley, Robert~B Penfold, Rebecca~C
  Rossom, Beth~E Waitzfelder, Katherine Sanchez, and Frances~L Lynch.
\newblock {Assessing and Minimizing Re-identification Risk in Research Data
  Derived from Health Care Records}.
\newblock {\em EGEMS (Washington, DC)}, 7(1):6, mar 2019.

\bibitem{tang2016protecting}
Haixu Tang, Xiaoqian Jiang, Xiaofeng Wang, Shuang Wang, Heidi Sofia, Dov Fox,
  Kristin Lauter, Bradley Malin, Amalio Telenti, Li~Xiong, et~al.
\newblock Protecting genomic data analytics in the cloud: state of the art and
  opportunities.
\newblock {\em BMC medical genomics}, 9(1):1--9, 2016.

\bibitem{trec}
Text {REtrieval} conference ({TREC}) home page.
\newblock \url{http://trec.nist.gov}, December 14 2001.

\bibitem{CensusHistory}
{U.S. Census Bureau}.
\newblock History of the decennial census.
\newblock
  https://www.census.gov/programs-surveys/decennial-census/about/history.html,
  April 2 2018.

\bibitem{fsrdc}
{U.S. Census Bureau}.
\newblock Federal statistical research data centers.
\newblock \url{https://www.census.gov/about/adrm/fsrdc.html}, 2021.

\bibitem{wang2017community}
Shuang Wang, Xiaoqian Jiang, Haixu Tang, Xiaofeng Wang, Diyue Bu, Knox Carey,
  Stephanie~OM Dyke, Dov Fox, Chao Jiang, Kristin Lauter, et~al.
\newblock A community effort to protect genomic data sharing, collaboration and
  outsourcing.
\newblock {\em NPJ genomic medicine}, 2(1):1--6, 2017.

\bibitem{WarrenBrandeis}
Samuel Warren and Louis Brandeis.
\newblock The right to privacy.
\newblock {\em Harvard Law Review}, 4(5), December 15 1890.

\bibitem{10.1145/1150402.1150499}
Raymond Chi-Wing Wong, Jiuyong Li, Ada Wai-Chee Fu, and Ke~Wang.
\newblock ($\alpha$, k)-anonymity: An enhanced k-anonymity model for privacy
  preserving data publishing.
\newblock In {\em Proceedings of the 12th ACM SIGKDD International Conference
  on Knowledge Discovery and Data Mining}, KDD '06, page 754–759, New York,
  NY, USA, 2006. Association for Computing Machinery.

\bibitem{xiao2020optimizing}
Yingtai Xiao, Zeyu Ding, Yuxin Wang, Danfeng Zhang, and Daniel Kifer.
\newblock Optimizing fitness-for-use of differentially private linear queries.
\newblock {\em arXiv preprint arXiv:2012.00135}, 2020.

\end{thebibliography}
\end{document}